**All-in-all-out magnetic domain size in pyrochlore iridate thin films as probed by local magnetotransport**

T. C. Fujita[1], M. Uchida[1,*], Y. Kozuka[1], S. Ogawa[1], A. Tsukazaki[2,3], T. Arima[4,5], and M. Kawasaki[1,5].

[1]*Department of Applied Physics and Quantum-Phase Electronics Center (QPEC), University of Tokyo, Tokyo 113-8656, Japan*

[2]*Institute for Materials Research, Tohoku University, Sendai 980-8577, Japan*

[3]*PRESTO, Japan Science and Technology Agency (JST), Tokyo 102-0075, Japan*

[4]*Department of Advanced Materials Science, University of Tokyo, Kashiwa 277-8561, Japan*

[5]*RIKEN Center for Emergent Matter Science (CEMS), Wako 351-0198, Japan*

[*]Correspondence should be addressed to uchida@ap.t.u-tokyo.ac.jp

**Abstract**

Pyrochlore iridates have attracted growing attention because of a theoretical prediction of a possible topological semimetal phase originating from all-in-all-out spin ordering. Related to the topological band structure, recent findings of the magnetic domain wall conduction have stimulated investigations of magnetic domain distribution in this system. Here, we investigate the size of magnetic domains in $Eu_2Ir_2O_7$ single crystalline thin films by magnetoresistance (MR) using microscale Hall bars. Two distinct magnetic domains of the all-in-all-out spin structure are known to exhibit linear MR but with opposite signs, which enables us to estimate the ratio of the two domains in the patterned channel. The linear MR for $80 \times 60$ $\mu m^2$ channel is nearly zero after zero-field cooling, suggesting random distribution of domains smaller than the channel size. In contrast, the wide distribution of the value of the linear MR is detected in $2 \times 2$ $\mu m^2$ channel, reflecting the detectable domain size depending on each cooling-cycle. Compared to simulation results, we estimate the average size of a single all-in-all-out magnetic domain as 1-2 $\mu m$.

Magnetic domain size and distribution are fundamental properties of magnetic materials, determined by a variety of competing energy scales including magnetostatic, domain wall, and anisotropy energies.[1] Controlling and detecting magnetic domains are prerequisites for spintronic device applications such as logic and memory devices employing domain-wall motion.[2–7] Due to the technological demand, magnetic domain structures have been intensively investigated in ferromagnetic materials using, for example, magneto-optical effect, x-ray circular dichroism,[8] and magnetic force microscopy.[9] Antiferromagnets are also an important class of materials to form a pinned layer in spin valve structures.[10] However, observing antiferromagnetic domains is challenging due to absence of the bulk net magnetic moment. While it has been enabled by various techniques,[11-17] the studies have been limited compared with the case of ferromagnetic materials.

Here we present an electrical detection of antiferromagnetic domains in thin films of a pyrochlore iridate $Eu_2Ir_2O_7$ using microscale Hall bars. $Ln_2Ir_2O_7$ ($Ln$: rare-earth elements) exhibits a paramagnetic-antiferromagnetic transition accompanied with a metal-insulator transition at $T_M$ depending on $Ln$ (Refs. 18,19). In the antiferromagnetic state, spins at $Ir^{4+}$ sites form all-in-all-out spin ordering, where all the four spins at the vertices of a tetrahedron pointing inward or outward and they are alternatingly stacked along <111> direction.[20–23] This magnetic ordering possesses two distinct spin arrangements all-in-all-out and all-out-all-in as shown in Fig. 1(a), which are the time-reversal counterparts each other. Hereafter, we refer to the two domains as A domain and B domain for simplicity. Previously, we reported that two distinct magnetic domains in $Eu_2Ir_2O_7$ thin films exhibit linear magnetoresistance (MR) of opposite signs.[24] This fact enables us to electrically distinguish the two domains and estimate their ratio within the thin-film channel.

Of particular interest in pyrochlore iridates is that metallic conduction is theoretically predicted along the antiferromagnetic domain wall as a result of topological nature of the band structure.[25,26] This metallic conduction is also experimentally observed, while the bulk state remains insulating in polycrystalline and singlecrystalline samples[27,28] and a heterostructure.[29] The creation and annihilation of domain walls are controllable by external magnetic field depending on $Ln^{3+}$ (Refs. 27,28,30), providing a great potential to open a new epoch of electronics. In order to utilize this domain wall state for future applications, it is necessary to investigate the magnetic domain size, since the device design largely depends on it. However, the study of the spatial distribution of all-in-all-out magnetic domains is limited so far. The domains have been very recently detected by x-ray circular dichroism in a bulk $Cd_2Os_2O_7$ single crystal[31] and scanning microwave spectroscopy in a bulk $Nd_2Ir_2O_7$ polycrystal.[32] On the other hand, investigation of the magnetic domains in thin films is highly important for opening the way for further investigations and applications of fascinating phenomena expected in this magnetic ordering. For example, carrier transport at single magnetic domain (or domain wall) can be characterized by using microfabricated devices on films thinner than magnetic domain size, and exchange bias can be induced by surface magnetic moment along <111> direction in heterostructures.

In this work, we evaluate the all-in-all-out domain size in $Eu_2Ir_2O_7$ thin films by magnetotransport measurement using microscale Hall bar structures down to $2 \times 2$ μm$^2$. This compound is an ideal platform for investigating the domains of all-in-all-out magnetic ordering of $Ir^{4+}$, since $Eu^{3+}$ has no magnetic moments. By measuring local magnetotransport within the microscale channel, we evaluate the spatial distribution ratio of A to B magnetic domains. Compared to simulation results, the average domain size is estimated as 1-2 μm from the statistical distributions of the value of linear MR.

$Eu_2Ir_2O_7$ films with a thickness of 50 nm were epitaxially grown on the Y-stabilized $ZrO_2$ (111) single crystalline substrate by combining pulsed laser deposition and solid phase epitaxy.[24] We used a mixed phase ceramics target with a prescribed ratio of Eu/Ir = 1/3 fabricated by a hot-press method at 950 °C under 25 MPa pressure. The films were deposited at a substrate temperature of 500 °C under an atmosphere of 100 mTorr Ar gas containing 1% $O_2$. The laser fluence and frequency were fixed at 6 J/cm$^2$ and 10 Hz, respectively. In order to probe the local magnetotransport, the sample was processed into microscale Hall bars by electron beam lithography. Ohmic Ni (10 nm) / Au (50 nm) contacts were deposited using an electron beam evaporator.

In Figs. 1(b) and 1(c), we show optical microscope images of a Hall bar with an electrode spacing of 2 μm and a width of 2 μm (2 × 2 μm$^2$ channel), where a pair of current electrodes ($I_+$, $I_-$) and 10 voltage electrodes are formed. As shown in Fig. 1(d), the longitudinal resistance ($R_{xx}$) exhibits a clear metal-insulator transition at $T_M$ = 105 K (Refs. 19,33). All the channels show qualitatively the same behavior, which indicates that the film is fairly homogeneous.

We then measured MR at 30 K after zero-field cooling (ZFC). The MR measurement was repeated at the same temperature after the sample was heated above 150 K, which is higher than $T_M$ = 105 K, and then cooled to 30 K to reset the magnetic domain structure. With repeating this measurement sequence, MR shows different linear components as exemplified in Figs. 2(a)–2(c), where the results after 1st, 4th, and 17th ZFC on the same channel (Ch4) are plotted. In this measurement, we did not find any appreciable hysteresis and thus show the data only for the downward field sweep throughout. Here we define the parameter $\alpha$ as

$$\text{MRR} = [R_{xx}(B) - R_{xx}(0\ \text{T})] / R_{xx}(0\ \text{T}) = (B\ \text{even term}) + \alpha B.$$

Characteristic of all-in-all-out spin ordering of $Eu_2Ir_2O_7$, we can assign the domain to A (B) if $\alpha$ is positive (negative).[24,34] Therefore, our observation demonstrates that the magnetic domain is randomly distributed and reconstructed in each cooling-cycle.

We also measured MR after field cooling under ±9 T for the same channels as shown in Fig. 2(d). As we have reported,[24,34] the domain structure in $Eu_2Ir_2O_7$ can be selectively stabilized to A or B single domain after field cooling above 3 T, showing MR with saturated gradients $\pm\alpha_s$. We thus regard the normalized value $\alpha_{\text{norm}} \equiv \alpha / \alpha_s$ for each channel as spatial distribution ratio of the A and B domains within the channel, i.e. $\alpha_{\text{norm}} = +1\ (-1)$ for A (B) single domain.

Figure 2(e) summarizes cooling-cycle dependence of $\alpha_{\text{norm}}$ for the $2 \times 2\ \mu m^2$ channel. $\alpha_{\text{norm}}$ shows random values centered at 0 for all the 6 channels, while $R_{xx}$ at 0 T remains at their original value during the measurements independent of cooling cycle (not shown). This fact assures that the observed change in $\alpha_{\text{norm}}$ is not induced by temporal degradation or spatial inhomogeneity of the magnetic domain structure. We performed the measurements described above for other Hall bars with different channel sizes in order to exclude the possibility that the domain size would be extrinsically determined by the channel size. In Fig. 2(f), we show $\alpha_{\text{norm}}$ of representative samples for $10 \times 10$ μm$^2$ and $80 \times 60$ μm$^2$ channels, revealing that the scattering away from $\alpha_{\text{norm}} = 0$ is more suppressed for larger channels.

In order to analyze the data quantitatively, we make probability histograms of $\alpha_{\text{norm}}$ between ±1 divided into 21 sections ($-1 \leq \alpha_{\text{norm}} < -0.95$, $-0.95 \leq \alpha_{\text{norm}} < -0.85$, $-0.85 \leq \alpha_{\text{norm}} < -0.75$, … , and $0.95 \leq \alpha_{\text{norm}} \leq 1$) as shown in Figs. 3(a)-3(c). The histograms show that $\alpha_{\text{norm}}$ is more widely distributed for the smaller channel. In the case of $10 \times 10$ μm$^2$ channel (Fig. 3(b)), $\alpha_{\text{norm}}$ is mostly concentrated around zero with a small fraction within $-0.15 \leq \alpha_{\text{norm}} < 0.15$. On the other hand, $\alpha_{\text{norm}}$ for the $2 \times 2$ μm$^2$ channel (Fig. 3(c)) is more widely distributed with tails, and the peak height of probability at $\alpha_{\text{norm}} = 0$ reduces to about 20%. This tendency is qualitatively consistent with a simple consideration that multi domains should be contained in a larger channel and the contribution to $\alpha$ from each domain should be cancelled out, while a smaller channel dominantly contains either detectable size of A or B domains. This result is also consistent with our previous report,[24] where $\alpha_{\text{norm}}$ shows only the zero value for the $80 \times 60$ μm$^2$ channel (Fig. 3(a)).

For a quantitative analysis, we perform a model simulation and estimate a ratio of the domain size (*D*) and channel size (*C*) by comparing the histograms. In this calculation, we adopt a simple model as follows: (i) A and B domains are modeled as squares with the length of *D*, and are assumed to be aligned in a checkerboard pattern. (ii) A square channel with the length of *C* is randomly located on that pattern with a rotational degree of freedom $\theta$. (iii) The areal contrast, (A−B) / (A+B), within the channel is assumed to be equal to $\alpha_{\text{norm}}$ (inset of Fig. 3(f)). We make histograms by repeating the steps (ii) and (iii) for 20,000 times with the ratio *D/C* ranging from 0.01 to 10. We confirm that the single domain magnetotransport ($\alpha_{\text{norm}} = \pm 1$) is not present for the case of *D/C* < 1.

Figure 3 compares the experimental and simulated histograms. Considering only that the histogram for the 80 × 60 channel is reproduced by the simulation for *D/C* < 0.1, we can only speculate that magnetic domain size is below about 10 μm. The experimental results for 10 × 10 and 2 × 2 μm$^2$ channels agree well with the simulations for *D/C* = 0.2 and 0.8, respectively. The histogram for 2 × 2 μm$^2$ channel does not show probability at $\alpha_{\text{norm}} = \pm 1$, which appears only for *D* >> 2 μm (*D/C* >> 1). On the other hand, $\alpha_{\text{norm}}$ should show much sharper peak centered at $\alpha_{\text{norm}} = 0$ if *D* << 1 μm (*D/C* << 0.5). Taking into account these facts, we can fairly estimate that the average domain size is 1-2 μm. In antiferromagnetic materials, large domain size is expected in general because no energy is gained by forming domain wall. In fact, a large domain size of ~ 50 μm has been reported for bulk $Cd_2Os_2O_7$ (Ref. 31) and ~ 5 μm for $Nd_2Ir_2O_7$ (Ref. 32), which also exhibit all-in-all-out spin ordering. The small domain size of our $Eu_2Ir_2O_7$ film is presumably attributed to internal strain or defects caused by lattice mismatch between the film and the substrate. A similar downsizing effect in the film has been also reported in NiO film (~ 50 nm)[13] compared with the bulk NiO (~sub mm).[35] Other effects such as stronger spin-orbit interaction of Ir or low dimensionality may be also effective, but further studies are necessary for ascertaining the origin.

To summarize, for estimating the all-in-all-out / all-out-all-in magnetic domain size in $Eu_2Ir_2O_7$ thin films, we have performed local magnetotransport measurement using microscale Hall bars. We have demonstrated that the linear MR after ZFC actually detects the spatial distribution ratio of the two distinct all-in-all-out magnetic domains contained in the channels. By comparing the experimental and calculated histograms of the linear term in MR, the average magnetic domain size is estimated as 1-2 μm. This study provides essential information about the domain size of pyrochlore iridate thin film, opening the way for detecting further fascinating transport phenomena predicted at the all-in-all-out magnetic domain wall.[27,32]

**Acknowledgments**

We thank J. Fujioka, K. Ueda, and Y. Tokura, for helpful discussions, and Y. Iwasa for the access to electron beam lithography equipment. This work was partly supported by Grant-in-Aids for Scientific Research (S) No. 24226002 and No. 24224010, by JSPS Fellowship No. 26·10112 (TCF), by Challenging Exploratory Research No. 26610098 (MU) from MEXT, Japan and by Casio Science Promotion Foundation (MU) as well as by Asahi Glass Foundation (YK).

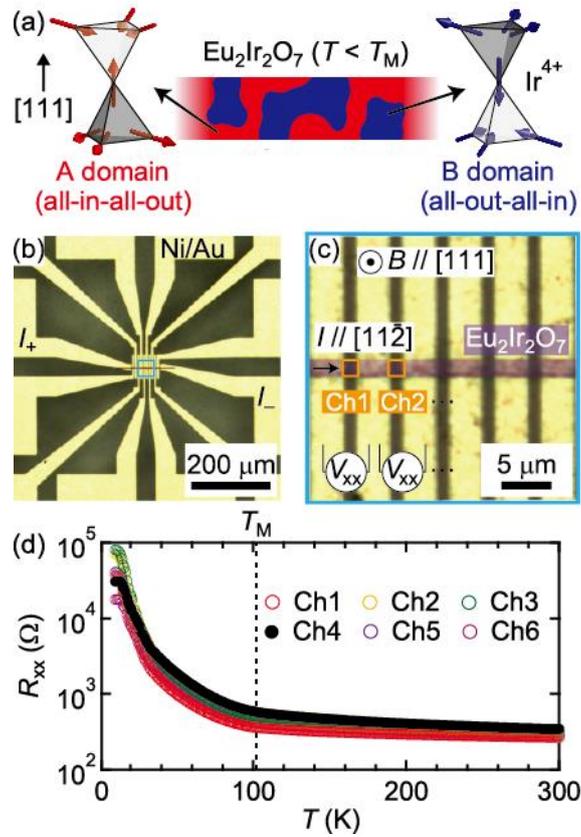

**Figure**

**Fig. 1** (a) Schematics of the all-in-all-out and all-out-all-in spin orderings at $Ir^{4+}$ site in pyrochlore lattice. Spatial distribution of these domains within a channel is also illustrated. (b) A microscope image of a Hall bar structure with $2 \times 2$ μm² channels. The area enclosed by the square at center is magnified in (c), where the film area is highlighted by false color in horizontal direction. The channels are located between Ni (10 nm) / Au (50 nm) electrodes and named as Ch1, Ch2, … . (d) Temperature dependence of the longitudinal resistance $R_{xx}$ for different 6 channels. The all-in-all-out or all-out-all-in magnetic ordering is accompanied by the metal-insulator transition at $T_M \sim 105$ K.

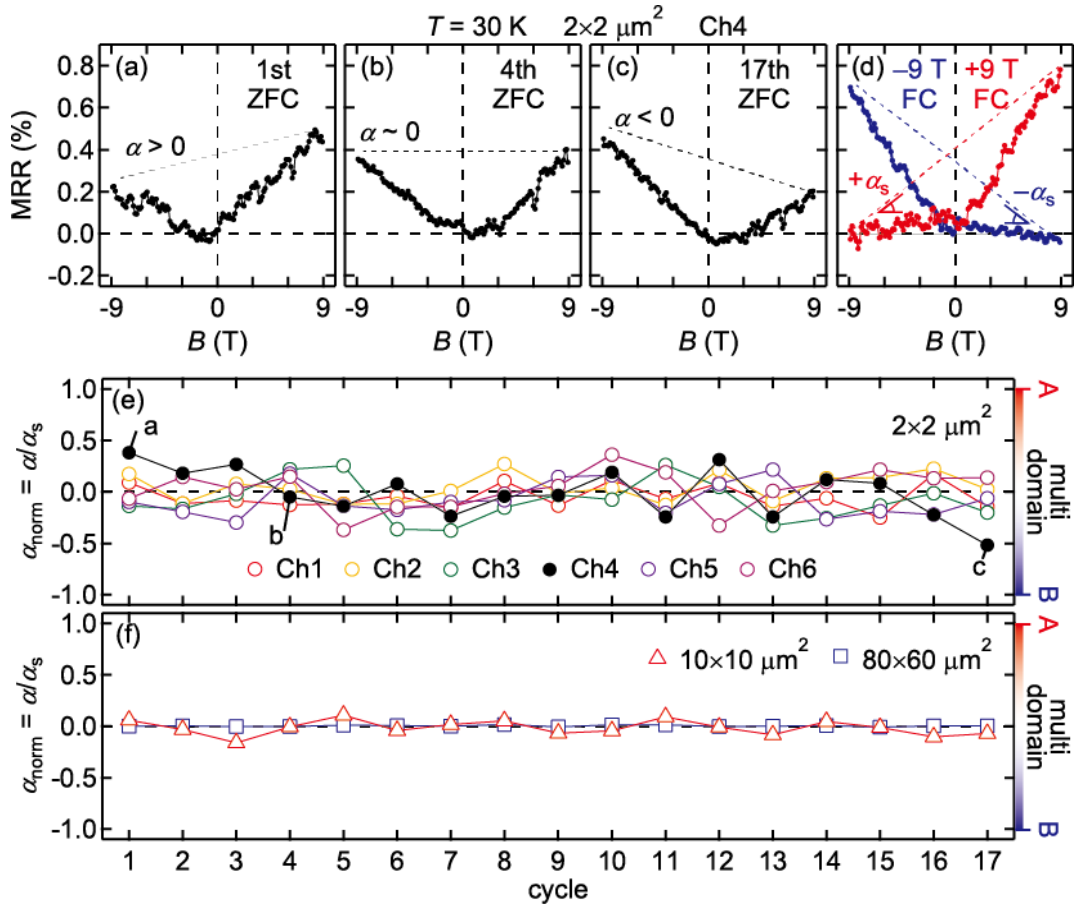

**Fig. 2** Magnetoresistance ratio (MRR) at 30 K for the Ch4 after 1st (a), 4th (b), and 17th (c) zero-field cooling (ZFC). The data after ±9 T field cooling (FC) are also shown in (d). The gradient of odd-parity MR is indicated by dashed lines, and termed as $\alpha$ for ZFC and $\pm\alpha_s$ for ±9 T FC. (e) Cycle-to-cycle variation of the normalized gradient $\alpha_{norm}$ ($\equiv \alpha / \alpha_s$) through 17 cycles for the different 6 channels. The data taken with the same sequence for representative channels of $10 \times 10$ and $80 \times 60$ μm$^2$ Hall bars are shown in (f).

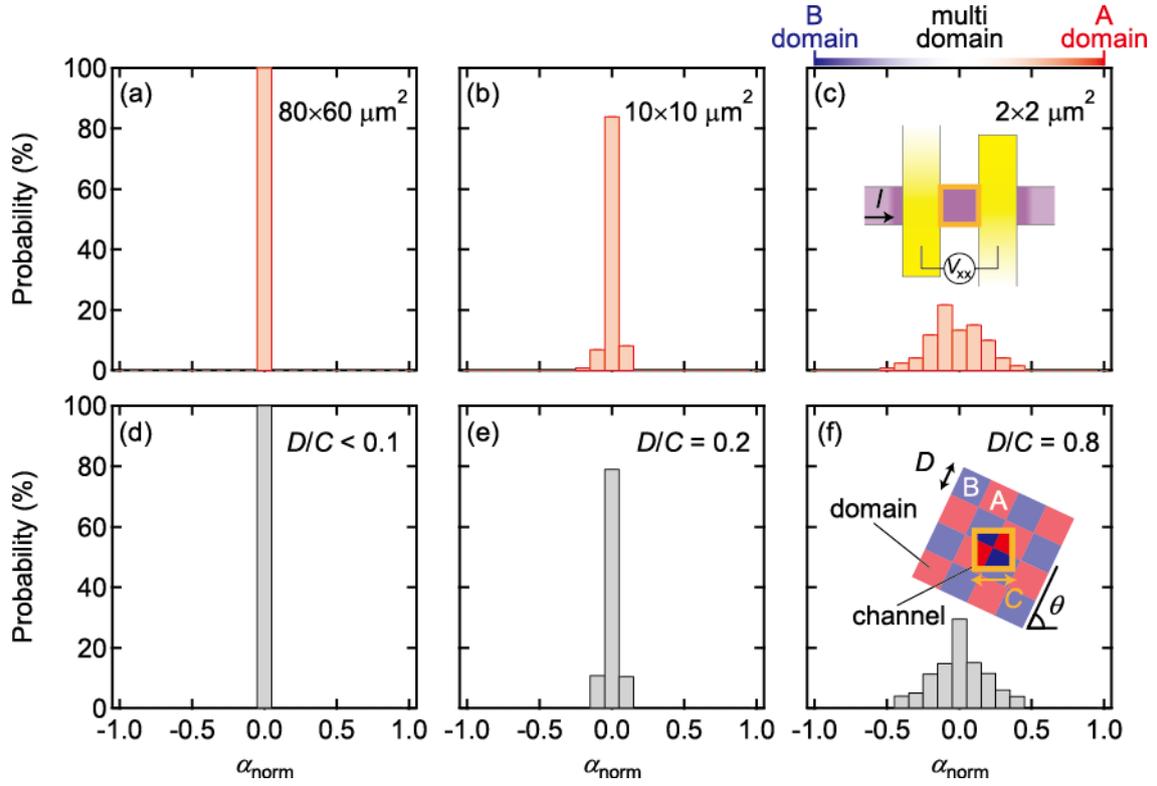

**Fig. 3** Probability histograms of $\alpha_{\text{norm}}$ experimentally deduced for (a) 80 × 60, (b) 10 × 10, and (c) 2 × 2 μm² channels. Inset in (c) schematically shows the measurement configuration. Calculated histograms of $\alpha_{\text{norm}}$ for various sets of the domain/channel size ratio (d) $D/C < 0.1$, (e) $D/C = 0.2$, and (f) $D/C = 0.8$. Inset in (f) illustrates a model used for the calculation. A square channel with a side length of $C$ (orange box) is randomly located at a rotational degree of freedom $\theta$ on A/B checkerboard domains with a length of $D$.